\documentclass[a4paper,11pt]{article}
\usepackage{aaskaiid}
\usepackage{orcidlink}

\title{Advancing extragalactic spectral line studies with the Square Kilometre Array Observatory}
\ShortTitle{Extragalactic Spectral Line}

\author[1]{Jacco Th. van Loon\orcidlink{0000-0002-1272-3017}}
\ShortName{van Loon et al.}
\author[2]{Viviana Casasola}
\author[3]{Manuela Bischetti\orcidlink{0000-0002-4314-021X}}
\author[4]{Sandra Etoka\orcidlink{0000-0003-3483-6212}}
\author[5]{Hans-Rainer Kl\"ockner \orcidlink{0000-0002-0648-2704}}
\author[6]{Mamta Pandey-Pommier\orcidlink{0000-0001-5829-1099}}
\author[7]{Mark Sargent\orcidlink{0000-0003-1033-9684}}
\author[8]{Nick Seymour\orcidlink{0000-0003-3506-5536}}
\author[9]{Andrea Tarchi\orcidlink{0000-0001-8540-3500}}

\affiliation[1]{Lennard-Jones Laboratories, Keele University, ST5 5BG, UK}
\emailAdd{j.t.van.loon@keele.ac.uk}
\affiliation[2]{INAF -- Istituto di Radioastronomia, Via P.\ Gobetti 101, 40129, Bologna, Italy}
\emailAdd{viviana.casasola@inaf.it}
\affiliation[3]{Dipartimento di Fisica "Enrico Fermi", Universit\`a di Pisa, Largo Bruno Pontecorvo 3, Pisa, I-56127, Italy}
\emailAdd{manuela.bischetti@unipi.it}
\affiliation[4]{Jodrell Bank Center for Astrophysics, University of Manchester, Oxford Road, Manchester M13 9PL, UK}
\emailAdd{sandra.etoka@googlemail.com}
\affiliation[5]{Max-Planck-Institut f\"ur Radioastronomie, Auf dem H\"ugel 69, D-53121 Bonn, Germany}
\emailAdd{hrk@mpifr-bonn.mpg.de}
\affiliation[6]{Pole Scientific, University Catholic of Lyon, Campus Saint-Paul, 10 place des Archives 69288, Lyon Cedex 02, France}
\emailAdd{mamtapommier@gmail.com}
\affiliation[7]{Institute of Physics, Laboratory of Astrophysics, \'Ecole Polytechnique Federale de Lausanne (EPFL), Observatoire de Sauverny, CH-1290 Versoix, Switzerland}
\emailAdd{mark.sargent@epfl.ch}
\affiliation[8]{International Centre for Radio Astronomy Research, Curtin University, GPO Box U1987, Bentley, WA 6845, Australia}
\emailAdd{nick.seymour@curtin.edu.au}
\affiliation[9]{INAF -- Osservatorio Astronomico di Cagliari, via della Scienza 5, 09047, Selargius (CA), Italy}
\emailAdd{andrea.tarchi@inaf.it}

\abstract{We present an overview of the Square Kilometre Array Observatory (SKAO) science potential in the area of extragalactic spectral lines besides $21$-cm neutral hydrogen. It highlights the main points from the SKAO Science Book chapters on individual topics, but is augmented by additional prospects. The SKAO will push studies and use of masers, kilomasers, megamasers, molecular and radio-recombination-line emission and absorption to a wider variety of environments, including to very high redshift where it may detect the first molecule (HeH$^+$). It will open the door to measurements of hydrogen and helium isotopes, probing the conditions for star formation and Big Bang nucleosynthesis. While the planned SKAO of the 2030s (AA*) is destined to forge major progress, an SKAO as initially envisaged (AA4) will truly transform the landscape, and an extension towards higher frequencies (up to $24$\,GHz) would enable water maser and ammonia surveys in nearby galaxies and systematic molecular gas inventories at redshift $5$.}

\begin{document}
\maketitle

\section{Preamble: what we mean by "extragalactic spectral line"}

The science working group "Extragalactic Spectral Line" was set up to complement the "H\,{\sc i} Galaxy Science" science working group. As such, H\,{\sc i} falls outside the remit of the extragalactic spectral line science working group, but obviously there is a lot of synergy both in terms of science and observing modes. In this overview chapter we shall focus on spectral lines other than H\,{\sc i}, but we do describe synergies with other observing modes including H\,{\sc i} at the end of this overview chapter. We also stress that H\,{\sc i} here means the $1420$\,MHz ($21$\,cm) spin-flip transition in neutral hydrogen; whilst hydrogen recombination lines are very much the realm of the extragalactic spectral line science working group.

Membership of the extragalactic spectral line science working group has doubled since the announcement of the $2025$ G\"orlitz meeting. In part, this is because astronomers have allowed themselves to dream of extending the science they have become accustomed to doing at cosmic noon ($z\sim2$) or in the nearby Universe all the way back to cosmic dawn, where the SKA will probe galaxies now known to exist at $z\sim14$. No less importantly, what was the exclusive domain of Galactic (Milky Way) science has gradually become possible in nearby galaxies of very different type and content, history or environment and the SKAO promises to fully develop this at scale across the nearby universe. While we claim anything that is not gravitationally bound to the Milky Way galaxy to be extragalactic we also include "foreign" objects that are traversing the galactic neighbourhood or have fallen prey to its gravitational clasp -- such as the SMC and LMC\footnote{We shall refer to these galaxies by their acronyms only as their commonly used full names bear associations with colonial heritage that we abhor.} that are also of interest to the "Cradle of Life" science working group, but we share with the "Our Galaxy" science working group embedded remnants such as the $\omega$\,Centauri globular cluster -- thought to be the nucleus of a disrupted dwarf galaxy.

\section{SKAO capabilities for extragalactic spectral line science}

In this chapter we summarise some of the unique science enabled by the expected (AA*) and aspired (AA4) capabilities, and in closing (\S~\ref{future}) outline in general how capabilities extending beyond the current deployment baseline capability (AA*) enable new science -- e.g., by virtue of higher sensitivity and angular resolution, or through a high-frequency extension (for more detail on the latter see \href{https://www.skao.int/sites/default/files/documents/d38-ScienceCase_band6_Feb2020.pdf}{SKA Memo 20-01 -- SKA1 Beyond $15$\,GHz: The Science case for Band 6}). In \S~\ref{future} we also discuss where synergies across SKAO interests or between SKAO and other flagship facilities enhances the scientific mission. The previous assessment a decade ago did not identify extragalactic spectral lines as a separate field \citep{bourke2015advancing}, so this is the first such assessment for SKAO capabilities \citep{braun2019anticipatedperformancesquarekilometre}.

\subsection{Sensitivity, resolution and survey speed}

With $144$ ($197$) dishes of $13.5$--$15$\,m diameter, the AA* (AA4) sensitivity of the SKAO is going to be transformational, but in many cases resolution matters too. Maser emission and cold gas are typically confined to angular scales of arcseconds or less, and these are the standard angular resolutions naturally obtained with SKA-Mid. Beyond AA2, zoom modes will allow the spectral resolution of $0.1$--$1$\,km\,s$^{-1}$ needed to kinematically and magnetically \citep{Robishaw01.2026.SKA} resolve masers and cold gas emission and absorption lines. Compared to the currently most powerful radio telescopes in the southern hemisphere that overlap in frequency range with the SKAO, the Australian SKA Pathfinder (ASKAP) and MeerKAT, this enables an order of magnitude increase in sensitivity and angular resolution at the same time (or pushing further in one, at the cost of the other).

The superior angular resolution matters, not only to be able to map detail in some cases but also to increase contrast, which aids detection, and cross-identification in multi-wavelength datasets. But despite this, SKA-Mid still offers a generous field-of-view, $\sim0\rlap{.}^\circ 6$ at $2$\,GHz, which makes it possible to map moderately extended objects such as nearby galaxies or star-forming regions in the LMC and SMC in a single pointing, or perform small sky surveys.

\subsection{Examples of indicative performance}

Given the broad scope of science requirements (see \S~\ref{science}) it is impossible to provide here a comprehensive overview of the SKAO capabilities for extragalactic spectral line work, but the table below offers an indication of what can be achieved across the SKA-Mid range. SKA-Low is mostly relevant to (stacked) radio recombination line studies which are described in \S~\ref{RRL}. Different choices of combinations of channel width (spectral resolution), synthesised beam size (angular resolution) and baseline weighting in image reconstruction (Briggs robust parameter $R$) for both the expected AA* and aspirational AA4 configurations are presented, at three different frequencies $\nu$ in bands 2 and 5 that correspond to the $z\sim0$ frequencies of hydroxyl (OH) and methanol (CH$_3$OH) but could represent redshifted lines of these or other transitions and species.

\begin{table}[h]
\caption{SKA-Mid indicative sensitivities$^1$ for spectral lines in bands 2, 5a and 5b. Pointing direction: SMC.}
\label{tab:SMCmasers}
\begin{tabular}{ccccccc}
\hline
 $\nu$ & Channels & Configuration & $R^2$ & Beam size$^3$ & rms noise & Integration time \\
 (GHz) & (km\,s$^{-1}$) & & & ($^{\prime\prime}$) & mJy\,beam$^{-1}$ & (hr) \\
\hline
$1.63^4$ & $\sim0.62$ & AA* & $+1$ & $2.87\times2.50$ & $1$ & $0.33$ \\
& $\sim0.15$ & AA* & $-1$ & $1.23\times0.96$ & $3$ & $1.33$ \\
& $\sim1.2$  & AA4 & $+1$ & $1.12\times0.84$ & $0.1^7$ & $7.5$ \\ 
& $\sim0.15$ & AA4 & $-1$ & $0.50\times0.35$ & $3$ & $0.42$ \\
\hline
$6.35^5$ & $\sim0.63$ & AA* & $+1$ & $0.57\times0.50$ & $1$ & $0.25$ \\
& $\sim0.08$ & AA* & $-1$ & $0.28\times0.20$ & $3$ & $2$ \\
& $\sim1.3$ & AA4 & $+1$ & $0.24\times0.18$ & $0.1$ & $5$ \\
& $\sim0.08$ & AA4 & $-1$ & $0.12\times0.08$ & $3$ & $0.42$ \\
\hline
$12.2^6$ & $\sim0.66$ & AA* & $+1$ & $0.30\times0.26$ & $1$ & $0.25$ \\
& $\sim0.08$ & AA* & $-1$ & $0.15\times0.10$ & $3$ & $2$ \\
& $\sim1.3$ & AA4 & $+1$ & $0.13\times0.09$ & $0.1$ & $4.5$ \\
& $\sim0.08$ & AA4 & $-1$ & $0.06\times0.04$ & $3$ & $0.42$ \\
\hline
\end{tabular}\\[2mm]
{\small
\noindent
1: from the sensitivity calculator: https://sensitivity-calculator.skao.int/mid.
2: Briggs robustness parameter. 
3: spectral synthesised angular resolution.
4: nominal frequency elected for the $1612$/$1665$/$1667$/$1720$\,MHz OH transitions.
5: nominal frequency elected for the $6.016$/$6.031$/$6.035$/$6.049$\,GHz OH \& $6.7$\,GHz CH$_3$OH transitions.
6: nominal frequency elected for the $12.2$\,GHz CH$_3$OH transition.
7: geared for the OH maser low-luminosity stellar population \citep{etoka2015}.}
\end{table}


\section{Summary of science book chapters and additional science topics}
\label{science}

\subsection{Radio recombination lines}
\label{RRL}

While the extragalactic spectral line work is often associated with SKA-Mid (or higher frequencies), SKA-Low is the prime observatory for Radio Recombination Lines (RRLs) as these transitions are strongest at low frequencies, but not exclusively by any means. The advantage of RRLs is that there are so many transitions related to the same species, that are regularly spaced in frequency and of which the relative intensities are easily predictable. This means that they can be stacked to boost sensitivity. The focus often is on hydrogen, for obvious reasons (there is a lot of it), and while strictly speaking this is neutral hydrogen (having recombined) we distinguish this from the $21$-cm line and its associated community and science working group even though the latter is also present in recombining gas (the warm neutral medium). Hydrogen RRLs have been detected out to redshift $z\sim1$ \citep{emig2019rrl1,emig2023rrl2}.

The SKAO potential for extragalactic RRL work is covered in \cite{Emig01.2026.SKA}, where it is also explained how RRLs from carbon are found in neutral gas, or even predominantly molecular gas, and that at the lowest frequencies they can dominate over the hydrogen RRLs that become suppressed by free--free absorption (which doesn't affect the neutral gas), while at high frequencies (upper end of SKA-Mid) spontaneous emission can become dominant (elsewhere, stimulated emission reigns and continuum sources bias towards high detection levels). That chapter also mentions helium RRLs, which is interesting since helium is chemically inert (though it can form helonium, HeH$^+$) and closely reflects Big Bang nucleosynthesis (BBN) -- though it has a higher ionisation potential. But helium RRLs seem still lying in wait to be fully exploited.

In addition, carbon RRLs can be detected in absorption against bright continuum sources, as they probe cold gas in front of hot gas. This opens new avenues towards studies of absorption line systems across a wide range of redshift, as well as in- or outflowing molecular gas in star-bursting or nuclear active galaxies. Exquisite kinematic detail can be obtained. Carbon RRLs were first detected outside the Milky Way in M\,82 by \citet{morabito2014rrl}, followed by Cas\,A \citep{salas2017rrl1,salas2018rrl2} and Cyg\,A \citep{cros2025rrl}. SKAO promises to open this extragalactic window.

\subsection{Megamasers}

Masers are amplified stimulated emission and can thus become extraordinarily bright. Their emission is beamed and the energy levels distribution is narrowed, making them excellent kinematic probes. They are also sensitive tracers of the specific conditions for population inversion and kinematic coherence to occur -- typically dense or shocked infrared-bright gas in regions of massive star formation and around evolved stars, or within the environments of Active Galactic Nuclei (AGN). The former are addressed in \S~\ref{masers}.

Masers in AGN, the 'so-called' megamasers, are discussed instead in \cite{Tarchi01.2026.SKA}, with a focus on water megamasers (H$_2$O, with a rest frequency at $22$\,GHz). These are unique tools to derive fundamental physical quantities of the host galaxies, such as precise black hole masses (from the disk rotation speed) and standard-candles-independent distances (from astrometry), and can provide clues on the interaction between nuclear jets/outflows and the interstellar medium (ISM).

\cite{Tarchi01.2026.SKA} provide a quantitative analysis of the expected number of new extragalactic water maser sources within reach of the SKA-Mid telescope (in AA4 configuration) through targeted and blind surveys, up to cosmological redshifts. In addition, the potential of high-spatial resolution follow-ups of the new sources is discussed, when Very-Large-Baseline Interferometry (VLBI) baselines are taken into consideration in the SKAO design.

While water megamasers provide a unique view of the central few parsecs of an AGN, hydroxyl (OH, with a rest frequency at $1.7$\,GHz) megamaser emission probes larger spatial scales and reveals a considerably more complex picture. Observations show both a diffuse, low-surface-brightness component extending over hundreds of parsecs and more compact maser emission regions associated with dust obscured structures in the central regions \citep{Baan2023,Kloeckner2003}. The diffuse component appears to be an effective tracer of outflows and jet interactions, whereas the compact emission is thought to arise from gravitational bound structures associated to AGNs and their dusty torus.

The hosts of the OH megamaser emission are generally believed to present a transient stage in galaxy evolution, during which a merger drives large quantities of gas and dust into the central region of a galaxy. This picture might be more complex and as shown in II\,Zw\,096, where a distinct OH megamaser has been detected far from the galactic nucleus \citep{Migenes2011,Wu2022}. Such observations suggest that the physical conditions required for OH megamaser activity may also arise outside the central nuclear environment. Determining whether OH megamaser emission is systematically linked to specific merger stages, however, requires substantially larger and statistically complete samples, complemented by multi-wavelength observations spanning the electromagnetic spectrum. These ancillary data are essential for characterising the host galaxies, constraining their merger state, and identifying the physical conditions that give rise to OH megamaser emission.

In this context, although no chapter in this book is explicitly and specifically devoted to the use of the SKAO for OH maser emission studies in AGN, there has been a recent resurgence in OH megamaser studies driven by the SKAO and its precursor/pathfinder telescopes both in conference contributions \citep[e.g.,][]{roberts2024oh} and serendipitous discoveries of OH megamasers at redshifts $z\sim0.5$--$1$ with the MeerKAT telescope \citep{glowacki2022oh,jarvis2024oh}. Even more recently, \citet{manamela2026oh} unveiled the most distant OH megamaser ever detected (redshift $z\approx1$) with a dedicated survey of gravitationally lensed targets, highlighting the possibility to exploit SKA-Mid and its precursors for OH megamaser surveys. The potential of OH megamaser surveys has been, however, mentioned in a number of chapters in this book as a 'commensal' product of H\,{\sc i} surveys \citep[e.g.,][]{Maccagni01.2026.SKA}.

\subsection{Circumstellar masers in nearby galaxies}
\label{masers}

Interferometric observations of masers in nearby galaxies at or below an arcsecond resolution increase detection contrast and aid cross-identification in multi-wavelength data, but in contrast to Galactic sources it will resolve extragalactic sources (with typical scales of $10$--$1000$\,au) only in exceptional cases. \cite{Rygl01.2026.SKA} covers circumstellar masers (cool evolved stars and star-forming regions), but largely focuses on the Milky Way.

\begin{figure}[h]
\centering
\includegraphics[width=1.0\columnwidth]{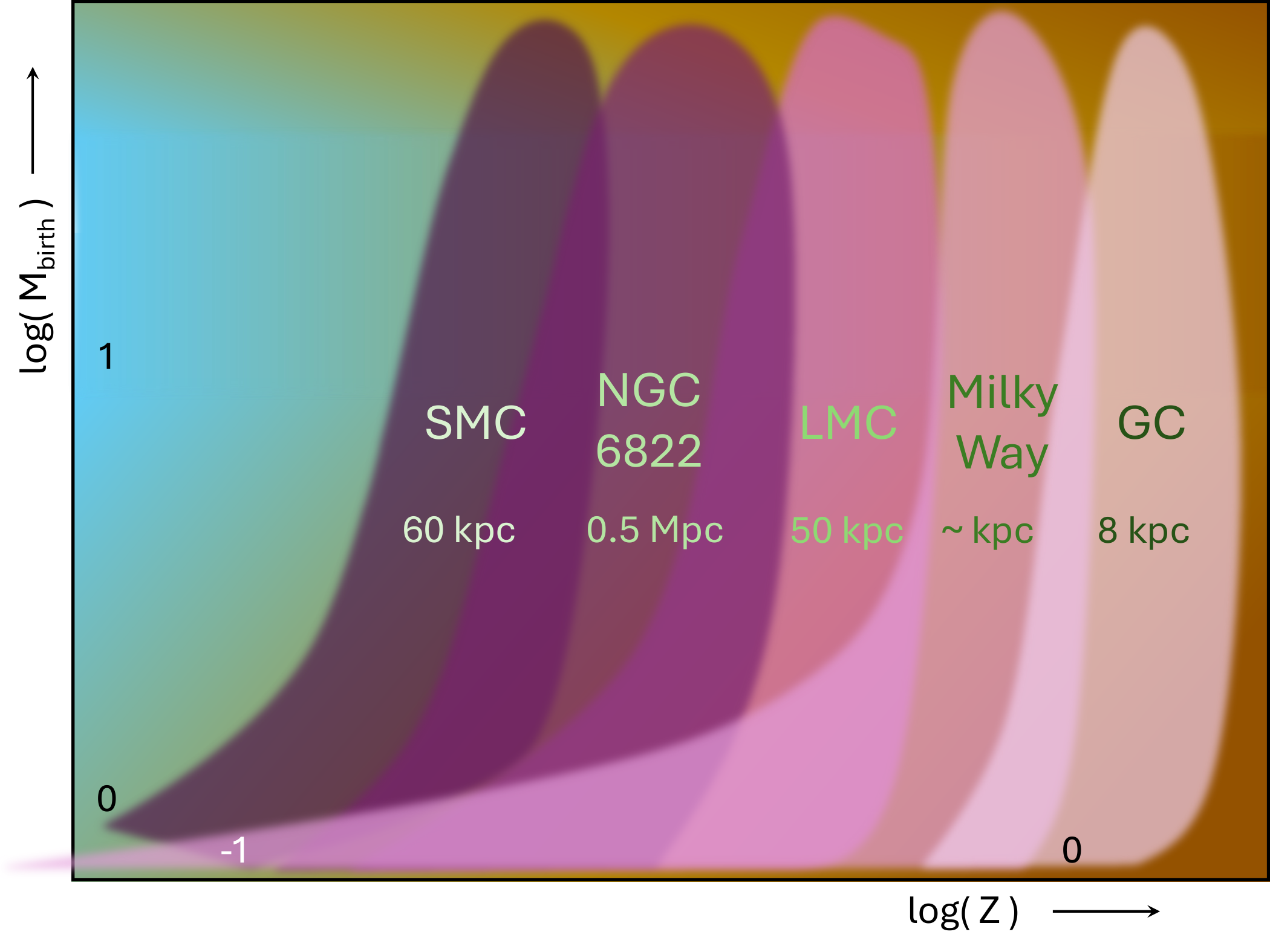}
\caption{Schematic overview of the birth masses and metallicities represented in the main stellar populations of the nearby galaxies LMC, SMC and NGC\,6822 compared to those in the Galactic Centre (GC) and the rest of the Milky Way. Figure courtesy J.\ van Loon.}
\label{fig:xgalpops}
\end{figure}

Advantages of studying evolved stars and star formation in nearby galaxies include the availability of different environments, primarily in terms of gas composition and star formation history and thus present-day stellar populations and their feedback. In particular, metal-poor high-mass star-forming gas dominates the LMC and SMC, at just $50$--$60$\,kpc distance to us, and NGC\,6822 about ten times further away, spanning a range in metallicity $\approx0.1$--$0.5$ solar, forming an excellent complement to similar populations in the Milky Way and its Galactic Centre at $\approx0.5$--$2$ solar metallicity (Fig.~\ref{fig:xgalpops}). These "large dwarf" galaxies can be studied in their entirety and they are well positioned for the SKAO.

Luminous, cool evolved stars are frequent hosts of OH masers (at $1612$, $1665$ and $1667$\,MHz in SKA-Mid band 2) that trace their slow, dense winds and directly measure the outflow dynamics. This is a unique probe of the mass-loss process, which is poorly understood but of great importance for the demise of the star and the regeneration of its surroundings \citep{vanloon2025massloss}. Extragalactic detections of such OH masers ($15$ in total) have so far been limited to the LMC \citep{wood1992oh,goldman2017lmc} but promising targets do exist in the SMC \citep{goldman2018smc}. Molecular gas shocked by an expanding supernova remnant can yield OH masers at $1720$\,MHz, which have indeed been detected in the LMC \citep{brogan2004oh1720}. Having only scratched the tip of the iceberg, the SKAO can uncover a substantial population of OH masers in the LMC, and has a high likelihood of detecting the first such masers in the SMC -- the most metal-poor ($\approx0.2$ solar) asymptotic giant branch stars and red supergiants for which such information would become available.

Masers allow unique access to the early feedback that regulates the star formation process, and that is important to chart in more pristine environments resembling the era of galaxy assembly from cosmic dawn to cosmic noon. They are also excellent beacons for the sites of protostars. Masers seen in such environments include excited OH at $6$\,GHz (band 5a) that probes cold dense gas, and tell-tale methanol masers (CH$_3$OH, at $6.7$ and $12.2$\,GHz in band 5a). These have been detected in the LMC but not the SMC \citep{green2008methanol,ellingsen2010methanol} nor in NGC\,6822 \citep{tarchi2020ngc6822} or M\,33 \citep{goldsmith2008triangulum}, but methanol maser emission was detected in M\,31 by \citet{sjouwerman2010andromeda}. Formaldehyde (H$_2$CO, at $4.8$\,GHz in band 5a and at $14.5$\,GHz in band 5b) is probably too weak in emission for large-scale extragalactic studies.

Zeeman splitting and polarisation (OH: circular; H$_2$O: linear) trace the magnetic field, and it will be possible to determine its relative importance in the formation and death of stars at low metallicity.

\subsection{Molecular absorption line studies}
\label{absorption}

Molecular absorption lines provide a powerful diagnostic tool for probing the radial motions of cold gas in both nearby and high-redshift galaxies. Unlike emission lines, which are biased toward high-density regions, absorption features depend linearly on the column density along the line of sight towards a bright background continuum source, allowing for the detection of low-density gas.

In nearby galaxies, observations of absorption features associated with the molecular gas (e.g., traced by OH or CO) have enabled the identification of non-circular motions, such as inflows and outflows \citep{gonzalezalfonso2017outflows,veilleux2020outflows}, identified by the velocity shift of the absorption relative to the systemic redshift. Studies at low redshift have reported the detection of molecular absorption in $\sim 20$\% of H\,{\sc i} absorbers, though this fraction is likely a lower limit due to the more severe dilution of molecular gas clouds in low-resolution observations compared to H\,{\sc i}. As an example, formaldehyde has been detected in absorption against galaxies ranging from the LMC to OH megamasers \citep{araya2004formaldehyde}, and a plethora of molecular absorption \citep{WiklindCombes} including OH (\citealp{CombesOH}; see also \citealp{Mahony01.2026.SKA}) and methanol \citep{Marshall2017} has been seen in a $z=0.89$ hydrogen-RRL emitting galaxy \citep{emig2023rrl2} against the backdrop of gravitationally lensed $z=2.5$ blazar PKS\,B1830$-$211.

Molecular absorption generally corresponds to the higher-velocity and broader H\,{\sc i} absorption component \citep[e.g.,][]{combes2024molecules}, effectively acting as a multi-scale probe: it traces the innermost ISM, where broad profiles reveal the kinematic impact of galactic outflows, as well as the larger scales of the circumgalactic medium. In these outer regions, the synergy between molecular and H\,{\sc i} absorption is crucial to identify accretion flows and outflows, providing a unique perspective on the mass exchange between galaxies and their environment.

In this context, the SKAO will provide the necessary sensitivity to conduct large-scale, systematic searches for molecular absorption lines, allowing for a quantitative assessment of gas accretion and feedback across a wide range of environments and cosmic epochs.

\subsection{Molecular emission line studies}

\cite{RanWang01.2026.SKA} describe the use of molecular lines that enter the SKA-Mid range at high redshift. These include low-$J$ (1--0) transitions principally of the key molecules CO, HCN and HCO$^+$ in cold star-forming gas, complementing high-$J$ transitions observed with the Atacama Large (sub-)Millimetre Array (ALMA) at higher frequencies. As a word of caution, low-$J$ CO transitions, especially CO(1--0), may be suppressed due to higher Cosmic Microwave Background (CMB) temperature at $z>6$ \citep{daCunha2013CMB} and warmer metal-poor molecular clouds \citep{vanloon2010warm,oliveira2011warm}. The SKAO potential for studying extragalactic spectral lines in gas-rich AGN hosting large-scale jets is covered in detail in \cite{Pandey-Pommier01.2026.SKA}.

\begin{figure}[h]
\centering
\includegraphics[angle=0,width=1.0\columnwidth]{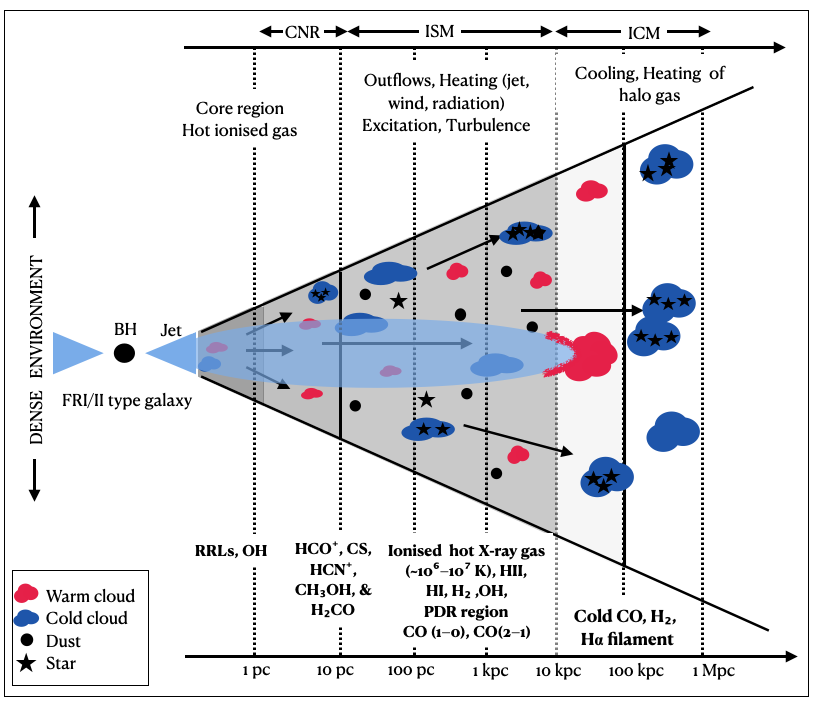}
\caption{Schematic diagram showing different regions and spatial scales over which AGN feedback operates. Multi-phase gas dynamics including molecular, atomic and ionised phases leads to formation of complex molecules. Adapted from \cite{Pandey-Pommier01.2026.SKA}.}
\label{fig:AGN-jet-gas}
\end{figure}

AGN feedback plays a fundamental r\^ole in regulating the cold gas reservoirs that fuel star formation and the growth of central black-hole systems in cosmic time. In both normal star-forming galaxies and radio-loud systems, feedback from AGN activity can drive turbulence, shocks, and multi-phase outflows that heat, redistribute, or deplete the cold gas required for star formation. These effects become even more pronounced in a sub-class of galaxies that tend to lie near dense cluster environments and that are rich in molecular gas such as H$_2$ and CO \citep{Ogle2010, Cicone2014,Tadhunter2014}. These galaxies host large reservoirs of molecular gas ($\sim2 \times 10^{10}$ M$_\odot$) together with powerful kpc--Mpc scale Fanaroff--Riley (FR)\,I/II radio jets and suppressed star-formation efficiencies. The presence of multi-phase gas clouds in the circumgalactic medium of these systems indicates that mechanical heating and turbulence can inhibit gravitational collapse despite the presence of abundant gas, as shown in Fig.~\ref{fig:AGN-jet-gas}. These processes can also trigger transitions between hot, warm, and cold gas phases through shock heating, turbulent mixing, and local thermal instabilities. While turbulent compression may promote the formation of denser molecular gas, the associated turbulent pressure and mechanical energy injection can prevent the gas from collapsing into self-gravitating star-forming clouds. 

Atomic and molecular spectral lines provide a powerful means to trace these feedback processes across different gas phases and physical scales. Neutral atomic gas has been traced through the H\,{\sc i} $21$-cm transition detected in emission up to $z\sim 0.4$ and in absorption out to $z\sim 3.37$ \citep{Kanekar2007, Fernandez2016}, while molecular gas reservoirs have been probed through low-$J$ transitions of molecules such as CO, OH, HCN, HCO$^+$, CH$_3$OH and H$_2$CO. In particular, molecular and H\,{\sc i} absorption lines are sensitive to diffuse and low-column-density gas (see \S~\ref{absorption}), enabling the detection of gas kinematics, and high-velocity outflows associated with AGN activity leading to jet-driven fragmented gas motions that may be difficult to detect in emission alone. Among OH megamasers, jet-host systems include galaxies such as Mrk\,231 ($z=0.032$) \citep{Reynolds2020,GonzalezAlfonso2014,Lonsdale2003}.

To quantify the amount of potentially star-forming gas and star formation efficiencies, we need to determine the mass in H$_2$, however this molecule is only visible under certain excitation conditions that do not resemble the bulk of the molecular gas. Instead, proxies are used, the most obvious of which is the second-most common, strongly-bound molecule CO. At high redshift, the low-$J$ transitions that more reliably trace the bulk of cold CO move into the SKA-Mid regime. CO emission has been detected out to $z\sim 6$ with ALMA and the Northern Extended Millimetre Array (NOEMA). [C\,{\sc ii}] observed with ALMA at high redshift complements but does not coincide with CO gas \citep{oliveira2019cii}, and neither does it trace the CO-dark molecular gas -- which is H$_2$ in which CO is undetectable. OH is haled as a potential tracer of CO-dark gas but its thermal emission is weak. The hope is that through a combination of tracers it will be possible to constrain the excitation conditions set by radiation and shocks, distinguishing nuclear feedback from intense star formation \citep{imanishi2026molecules}.

Molecular gas disks of galaxies may be resolved at $0\rlap{.}^{\prime\prime}1$ scales, providing kinematic maps, even at high redshift. The cold molecular component of galaxy outflows can also be quantified, which is crucial for understanding quenching and the enrichment of the intergalactic medium. SKA-Mid will struggle in band 5b at $z>6.5$, but star-burst galaxies ought to be detectable. Sub-mm/infrared-luminous galaxies, biased in massive halos, and gravitationally-lensed star-forming galaxies would be prime targets.

\subsection{Non 21-cm hyperfine atomic transitions}

Not discussed elsewhere in this book, there are other hyperfine atomic transitions in addition to the classic $21$-cm line which are potentially observable in extragalactic sources. Here we highlight the two most likely candidates: the ground states of deuterium (a hydrogen atom with one neutron) and the helium-3 ion (an ionised helium atom with only one neutron). These are the most common isotopologue variants of elements lighter than carbon with hyperfine transitions at radio wavelengths \citep[see][for other hyperfine transitions from light elements]{Karshenboim2002}.

Deuterium \citep[$\nu_{\rm hyperfine}=327.384\,$ MHz;][]{nafe1948} has an abundance about $40,000$ times less than hydrogen \citep{Coc2014}. Deuterium is nearly exclusively formed in Big Bang Nucleosynthesis (BBN) hence its abundance today is very close to its primordial value. This value was fixed around twenty minutes after the Big Bang; although deuterium is destroyed in stellar fusion, its fraction in pristine gas should be unchanged. As such its relative abundance to hydrogen is a key constraint on conditions on the very early Universe, in particular the baryon density. At present the deuterium primordial abundance is best constrained by the Planck constraints on the baryon density \citep{planck2020} rather than direct observation.

Deuterium is also detectable via its recombination lines in the ultraviolet and optical \citep[in the ISM, H\,{\sc ii} regions and distant quasars;][respectively]{Linsky2006,Hebrard2000,Burles1998} as well as from the hyperfine lines of deuterated molecules in interstellar clouds \citep[e.g.,][]{wilson1973}. However, the $327$\,MHz hyperfine transition provides the potential to detect, with SKA-Low, atomic deuterium, D\,{\sc i}, in pristine cold molecular clouds. The use of this line to detect interstellar deuterium was first proposed by \citet{Shklovsky1952} and immediately attempts to detect this line were made \citep[e.g.,][]{getmanzev1957,adgie1957,stanley1956}. Among other non-detections in subsequent years, a marginal detection of absorption of the 327\,MHz line towards the Galactic centre was presented by \cite{Cesarsky1973}. A $3$-$\sigma$ detection of the $327$\,MHz emission line toward the Galactic anti-centre was reported by \citet{rogers2007}. Notable potential targets for the SKAO are nearby undisturbed galaxies with bright H\,{\sc i} emission \citep[e.g., from the HIPASS survey;][]{Koribalski2004} which could potentially provide a precise measure (or at the very least a good upper limit) of the primordial deuterium abundance independent of CMB observations or distant quasars.

It should be possible to detect D\,{\sc i} in the LMC and SMC, especially at the highest column densities where H\,{\sc i} becomes optically thick -- brightness temperatures around 100\,K are typical so mK sensitivity suffices. This will help determine spin temperatures and cold (molecular) gas fractions. This is important as it characterises the conditions in star-forming gas, and is likely to lead to improved understanding of the transition from atomic gas through molecular gas to stellar plasma.

Singularly ionised helium-3 ($^3$He$^+$, $\nu_{\rm hyperfine}=8.666\,$\,GHz) was first proposed as a potential astrophysically observable feature by \cite{townes1957}. It was then first detected in giant H\,{\sc ii} regions \citep{rood1979}. Helium-4 too is originally formed in BBN, making up $\sim 8$\% of the number of atoms, but $\sim 25$\% of the mass, of the Universe. It is also produced in stellar nucleosynthesis, hence its abundance is expected to be higher in regions with substantial stellar processing. Helium-3 has a substantially lower primordial abundance than helium-4 \citep[by about $100,000$ times less than hydrogen;][]{Kneller2004} making any signal much weaker. However, its shorter hyperfine line wavelength compared to deuterium and its larger spontaneous decay rate compared to H\,{\sc i} (by a factor $\sim 680$) has made it an attractive prospect to astronomers. Galactic observations of $^3$He$^+$ have been used to measure its current abundance \citep[e.g.,][]{bania2010}, but deriving a primordial abundance is not straightforward due to the difficulty of determining its stellar production and destruction \citep[see][for a discussion]{Olive1995}. Given its high frequency, observations of the $8.666$\,GHz line up to $z\sim 0.9$ can be performed in band 5.

While not yet detected outside the Galaxy, the $^3$He$^+$ line is also thought to be a good measure of the final stages of helium re-ionisation. Helium gets ionised in two stages; the first electron has a similar binding energy as that of the hydrogen atom so gets removed during hydrogen re-ionisation ($z>6$) leaving singularly ionised helium-4 (and helium-3). The second electron has a much higher binding energy and requires a more intense ultraviolet radiation field to be ionised -- likely provided by AGN. Helium ionisation is likely complete by $z\sim 3$ \citep{Furlanetto2008}. Hence, $^3$He$^+$ signatures of helium reionisation should be detectable over $0.9<z<5.1$. Above $z\sim5.1$ the observed-frame signal would overlap with that from local and moderately-redshifted H\,{\sc i}, complicating detection and interpretation of the $^3$He$^+$ line.

The $^3$He$^+$ could potentially be detected directly via absorption of the hyperfine line against compact $z>3$ radio galaxies. While rare, luminous radio sources at $z\ge 3$ are being found \citep[e.g.,][]{seymour2024} offering the chance to detect $^3$He$^+$ absorption lines in band 2 (or 3 when available). While predicted theoretically \citep{bagla2009} the only attempt to directly measure helium reionisation via the cross-correlation power spectrum of the redshifted hyperfine line was by \cite{trott2024}. This work was more a proof of principle as the sensitivity is many orders of magnitude too shallow. Predictions for detections by SKA-Mid are given in \cite{spina2026}.

\subsection{The first cosmic molecule -- HeH$^+$}

The first molecule to form during the recombinaton era some 400,000 years after the Big Bang was protonated helium \citep{HeH+2013}: the helium hydride cation "helonium" (HeH$^+$). Its neutral form, HeH is unstable in its ground state.

While the sole detection up to now has been made in the young, dense planetary nebula NGC\,7027 \citep{HeH+2019}, detecting it around or before the epoch of reionisation would unlock a new way to characterise the conditions in the Universe when it switched on its lights. Attempts to detect it against a $z=6.4$ quasar have been made, but failed \citep{HeH+2011}.

The SKAO offers the prospect of detecting HeH$^+$ at cosmological distances. The rest wavelength of $149.14$\,$\mu$m is shifted into the SKA-Mid range at $z>130$. This coincides with an important transition in cosmic history as matter thermally decoupled from the CMB, cooling more rapidly. SKA-Mid can trace it all the way back to the recombination epoch at $z\sim1100$ where HeH$^+$ appears at a frequency of $1.8$\,GHz. We encourage observational cosmologists to start thinking of ways in which HeH$^+$ might be detected at such early times.

\section{Synergies with other SKA modes and other facilities}
\label{future}

\subsection{Other SKA modes, planned and unplanned}

Already among different types of spectral line studies we have demonstrated strong synergies when it comes to tracing multiple gas phases in nearby and distant galaxies, in accretion and outflows. Naturally, there is also a strong connection to H\,{\sc i} \citep{JingWang01.2026.SKA}, which can be detected in absorption in similar sources where molecules or RRLs are seen. H\,{\sc i} surveys may produce serendipitous detections of redshifted OH, or RRLs -- especially when they extend to high redshift.

Obviously, line-absorption studies require discrete continuum background sources (if not the Cosmic Microwave Background). The ionised ISM or jets traced by radio continuum emission also provide the physical context for interpreting atomic and molecular line emission and absorption. Likewise, polarisation constrains the r\^ole of magnetic fields. High-redshift intensity-mapping experiments for [C\,{\sc ii}] or CO lines would of course also be highly relevant to extragalactic spectral line work.

The interferometric capabilities of the SKAO already benefit studies of small-scale phenomena such as cold gas clumps or masers, or high-redshift source detection. VLBI is an additional mode of particular interest to the use of megamasers for calibrating the cosmic distance ladder.

Astronomical maser variability, while common, has not been discussed here, but the long-term operational capabilities and exceptional system stability of SKA-Mid will facilitate unprecedented monitoring of extragalactic (and Galactic) maser sources. These observations will offer unique insights into the maser mechanism, the underlying physical conditions of the emitting regions, and the evolutionary and dynamical processes taking place within environments that are otherwise inaccessible to direct observation.

The above science cases make it clear that in many cases first detections, first statistically meaningful analysis or extension towards higher redshift become possible with the planned AA* configuration, but improvement in sensitivity provided by a full AA4 implementation would make a significant difference in the reliability, detail, statistics and success with which these studies can be carried out.

Our science working group strongly advocates for a band 6, extending the SKAO capabilities to frequencies $>15$\,GHz, up to $\sim 24$\,GHz. Water masers (at $22$\,GHz) have been detected around evolved stars \citep{vanloon2001lmcEvolved} and more often in young stellar objects in the LMC \citep{vanloon2001lmcYSO,oliveira2006lmc,imai2013lmc}, and in the latter case also in the SMC \citep{breen2013smc}, M\,31 \citep{darling2011andromeda}, M\,33 and IC\,10 \citep{brunthaler2006triangulum} and in star-bursting galaxies (at high luminosities, traditionally labeled `kilomasers') such as the Antenn{\ae} at $\sim20$\,Mpc distance \citep{brogan2010antennae} and the merging system (and megamaser host) Arp\,299 at $\sim42$\,Mpc distance \citep{tarchi2011arp299}, though not in NGC\,6822, IC\,1613 or WLM \citep{tarchi2020ngc6822}. Water masers probe different regions in the outflows from hydroxyl and methanol, thus completing the energy and kinematic picture. Ammonia (NH$_3$ (1,1) and (2,2) inversion transitions around $23.7$\,GHz) is an excellent "thermometre" of cold dense gas, and has been detected in the LMC \citep{ott2010methanol} but not SMC \citep{lee2009methanol}. Band 6 would substantially improve access to redshifted low-$J$ molecular transitions such as CO(1--0), HCO$^+$ and methanol during the peak epoch of galaxy assembly ($z\sim4$), narrowing the gap in redshift space between ALMA and SKAO.

\subsection{Other facilities, online and upcoming}

The Karl G.\ Jansky Very Large Array (VLA) -- especially after its upgrade in the early 2010s -- remains a primary workhorse for cm-wave science in the Northern Hemisphere, providing essential access to low-$J$ CO transitions, OH absorption and maser transitions that complement SKA-Mid. The next-generation VLA (ngVLA) is expected to become the ideal high-frequency partner, operating up to $116$\,GHz and bridging the gap between the SKAO's decimetre capabilities and the millimetre regime. This is particularly relevant for high-resolution studies of nuclear environments and dense gas tracers. At higher frequencies, ALMA and its ALMA2040 roadmap remains the premier facility for probing the dusty ISM and high-$J$ molecular transitions. Its wide-band sensitivity upgrade provides the bandwidth necessary to follow up SKAO detections of RRLs and varied maser species -- its band 2 frequency domain now extends downwards to $67$\,GHz, giving access to the $86$\,GHz SiO transitions in addition to those at $43$\,GHz SiO in ALMA band 1 ($35$--$50$\,GHz).

The combination of next-generation radio observatories enables powerful VLBI opportunities; we refer to \cite{Radcliffe01.2026.SKA} for a more thorough discussion.

The diagnostic power of extragalactic radio spectral lines is maximally leveraged when integrated into a multi-wavelength framework. The spaceborne James Webb Space Telescope (JWST) provides the star formation context and ionised gas kinematics, while the upcoming groundbased $30$--$40$-m class telescopes will resolve stellar populations and nuclear environments at spatial scales matching the SKA-Mid spectral line maps.

Together, these facilities ensure a holistic view of galaxy assembly and evolution, black hole accretion, and the physical laws governing the universe across cosmic time.

\section{Final statement}

Astronomers have come to realise the great potential to combine multi-phase gas tracers to gain a complete picture of the physical, chemical and dynamical state of the gas that feeds star formation across cosmic time and that has ultimately resulted in planet Earth and humanity and all other life. Such science is not done within one working group, nor with one facility and certainly not by one community of scientists that are privileged to have access to resources and visibility. We therefore wholeheartedly endorse actions that lead to greater inclusivity and equity-based cooperation across political and socio-economical boundaries, for the benefit of all.

\bibliographystyle{abbrvnat-maxbibnames4}
\newcommand{\actaa}{Acta Astron.} 
\newcommand{\araa}{ARA\&A} 
\newcommand{\aar}{A\&ARv} 
\newcommand{\aapr}{A\&ARv} 
\newcommand{\ab}{Astrobiol.} 
\newcommand{\aj}{AJ} 
\newcommand{\apj}{ApJ} 
\newcommand{\apjl}{ApJL} 
\newcommand{\apjs}{ApJSS} 
\newcommand{\ao}{Appl. Opt.} 
\newcommand{\apss}{Astro. \& Space Sci.} 
\newcommand{\aap}{A\&A} 
\newcommand{\aaps}{A\&AS.} 
\newcommand{\baas}{Bull. Am. Astron. Soc.} 
\newcommand{\caa}{Chinese A\&A} 
\newcommand{\cjaa}{Chinese J. A\&A} 
\newcommand{\cqg}{Class. Quantum Gravity} 
\newcommand{\gal}{Galaxies} 
\newcommand{\gca}{Geo. Cosmo. Acta} 
\newcommand{\icarus}{Icarus} 
\newcommand{\jcap}{JCAP} 
\newcommand{\jgr}{J. Geophys. Res.} 
\newcommand{\jgrp}{J. Geophys. Res. Planets} 
\newcommand{\jqsrt}{J. Quant. Spectrosc. Radiat. Transf.} 
\newcommand{\memsai}{Mem. SAIt} 
\newcommand{\mnras}{MNRAS} 
\newcommand{\nat}{Nature} 
\newcommand{\nastro}{Nat. Astron.} 
\newcommand{\ncomms}{Nat. Commun.} 
\newcommand{\nphys}{Nat. Phys.} 
\newcommand{\na}{New Astron.} 
\newcommand{\nar}{New Astron. Rev.} 
\newcommand{\physrep}{Phys. Rep.} 
\newcommand{\pra}{Phys. Rev. A} 
\newcommand{\prb}{Phys. Rev. B} 
\newcommand{\prc}{Phys. Rev. C} 
\newcommand{\prd}{Phys. Rev. D} 
\newcommand{\pre}{Phys. Rev. E} 
\newcommand{\prx}{Phys. Rev. X} 
\newcommand{\prl}{Phys. Rev. Let.} 
\newcommand{\psj}{Planet. Sci. J.} 
\newcommand{\planss}{Planet. Space Sci.} 
\newcommand{\pnas}{Proc. Natl Acad. Sci. USA} 
\newcommand{\procspie}{Proc. SPIE} 
\newcommand{\pasa}{PASA} 
\newcommand{\pasj}{PASJ} 
\newcommand{\pasp}{PASP} 
\newcommand{\rmxaa}{RMXAA} 
\newcommand{\sci}{Science} 
\newcommand{\sciadv}{Sci. Adv.} 
\newcommand{\solphys}{Sol. Phys.} 
\newcommand{\sovast}{Soviet Ast.} 
\newcommand{\ssr}{Space Sci. Rev.} 
\newcommand{\uni}{Universe} 

\setlength{\bibsep}{0.0pt}
\bibliography{xGalLine}

\end{document}